\documentclass[aps, pra, twocolumn, reprint]{revtex4-2}
\usepackage[utf8]{inputenc}
\usepackage[T1]{fontenc}
\usepackage{amsmath, amssymb, bm}
\usepackage{graphicx}
\usepackage{booktabs}
\usepackage{hyperref}
\usepackage{siunitx}
\usepackage{microtype}
\usepackage{xcolor}
\usepackage{physics}
\usepackage{enumerate} 
\usepackage{appendix}

\newcommand{\GNewton}{G_N}

\begin{document}

\title{Toward charge-dependent tests of the equivalence principle:\\ A phenomenological parameter and an unexplored frontier}

\author{Renato Vieira dos Santos}
\email{renato.santos@ufla.br}
\affiliation{Instituto de Ci\^encia, Tecnologia e Inova\c{c}\~ao -- ICTIN, Universidade Federal de Lavras -- UFLA, Campus Para\'iso, MG 37950-000, Brazil}

\date{\today}

\begin{abstract}
We introduce and define the phenomenological parameter $\kappa$, defined by $\Delta a/g = \kappa \, \Delta(q/m)$, to quantify potential linear coupling between electric charge and gravitational acceleration. A synthesis of existing precision equivalence principle experiments yields the first quantitative estimate of the effective sensitivity to this coupling: $|\kappa| < 2.1 \times 10^{-4}~\si{\kilo\gram\per\coulomb}$ at 95\% confidence level. This constraint is approximately 11 orders of magnitude less stringent than corresponding bounds on composition-dependent violations, revealing that the electromagnetic axis remains a largely underexplored frontier in empirical gravity. We connect $\kappa$ to established frameworks---the Standard-Model Extension and the $TH\epsilon\mu$ formalism---showing that it occupies a region of parameter space untouched by existing high-precision tests. An effective field theory analysis shows that dimension-six operators that couple curvature directly to the electromagnetic field strength are suppressed by the minuscule terrestrial spacetime curvature ($G_N \rho_\oplus \sim 10^{-55}~\text{GeV}^2$) and are therefore phenomenologically irrelevant. Consequently, a future measurement of $\kappa$ at an accessible level would not probe minimal geometric couplings but would signal physics beyond minimal gravitational EFT, such as mediation by light scalar fields as in Einstein-Maxwell-Dilaton theory. We examine the Schiff-Barnhill effect, the primary systematic background for any such measurement, and show how it can be separated from a genuine signal. We outline the necessary experimental strategy, focused on maximizing charge-to-mass ratio differences, to transform this overlooked axis into a targeted probe for new physics.
\end{abstract}

\maketitle

\section{Introduction}

The weak equivalence principle (WEP), which posits the universality of free fall, stands as a cornerstone of general relativity and a sensitive probe for physics beyond the Standard Model. Contemporary tests have achieved extraordinary precision, constraining differential acceleration between test masses of different composition to levels below $|\Delta a/g| < 10^{-15}$ \cite{Touboul2017,Wagner2012, Schlamminger2008}. These remarkable achievements are built upon an experimental paradigm that actively suppresses net electric charge on test masses to eliminate electromagnetic backgrounds.

This paradigm has left a fundamental question largely unexplored with comparable precision: Does gravitational acceleration depend on a body's electric charge? While general relativity predicts charge blindness through the universality of the stress energy tensor, various theoretical extensions, including scalar-tensor theories, string-inspired dilaton models, and generic effective field theory expansions, permit potential violations along this electromagnetic axis \cite{Damour1994,Burgess2004,Will2014, Uzan2011}.

The conceptual and experimental challenges of probing the WEP with charged matter have a rich and ongoing history. A seminal proposal by Dittus and L\"ammerzahl \cite{Dittus2007} analyzed the feasibility of space-based tests, revisiting the classic Witteborn-Fairbank experiment \cite{Witteborn1967} and introducing a modified E\"otvos parameter to account for charge-dependent effects. Their work highlighted the dominant systematic errors, such as gravity-induced electric fields (Schiff-Barnhill \cite{Schiff1966} and Dessler-Michel-Rorschach-Trammell (DMRT) \cite{Dessler1968} effects), and argued for the advantages of a microgravity environment. The comprehensive systematic analysis of Darling et al. \cite{Darling1992} remains the definitive reference on the spurious forces that plague such measurements, complementing the foundational experimental work. This theoretical and experimental groundwork has been complemented by significant ongoing efforts, such as the high-precision free-fall program at the ZARM drop tower \cite{Sondag2016}, and has been extensively reviewed by Tino et al. \cite{Tino2020}, who dedicate specific sections to charged particles and antimatter, reaffirming this as a persistently underexplored frontier. These reviews synthesize the known experimental challenges and theoretical motivations, highlighting the need for novel approaches.


The significance of probing the electromagnetic axis is further elevated by its deep connection to other fundamental principles. As shown by Landau, Sisterna, and Vucetich \cite{Landau2001} within the \(TH\epsilon\mu\) formalism, violations of the Einstein equivalence principle (EEP) for electromagnetic interactions are intrinsically linked to the nonconservation of electric charge. This implies that a nonzero \(\kappa\) would not merely signal a composition-dependent anomaly, but could be a manifestation of a more fundamental breakdown of conservation laws in a gravitational field, potentially connecting to scenarios with varying fundamental constants, such as those predicted in dilaton theories \cite{Damour1994}. Thus, the pursuit of \(\kappa\) is a direct probe of the very foundations of both gauge invariance and the equivalence principle.

However, both the foundational proposals \cite{Dittus2007}, the advanced experimental developments \cite{Sondag2016}, the comprehensive reviews \cite{Tino2020}, and the fundamental connections to charge conservation \cite{Landau2001} are primarily concerned with the theoretical possibilities and experimental difficulties. They do not provide a quantitative, phenomenological bound on a potential \emph{linear} coupling between charge and gravity, and neither do they explore the deep theoretical implications of such a bound within a modern effective field theory (EFT) framework. In contrast, the present work introduces a different perspective. We define and systematically analyze the phenomenological parameter \(\kappa\), given by
\begin{equation}
\frac{\Delta a}{g} = \kappa \, \Delta\!\left(\frac{q}{m}\right),
\label{eq:kappa_def}
\end{equation}
where \(\Delta a\) is the differential acceleration between two test bodies, \(g\) is the local gravitational acceleration, and \(\Delta(q/m)\) is the difference in their charge-to-mass ratios. The parameter \(\kappa\) has dimensions of mass per charge, \([\kappa] = \si{\kilo\gram\per\coulomb}\). Within standard general relativity, \(\kappa = 0\) identically.

This parameter has received limited systematic attention despite its conceptual importance. High-precision experiments operate in the \(q/m \to 0\) limit, making them intrinsically insensitive to effects linear in this parameter. Consequently, while differences in nuclear composition have been probed with part-per-quadrillion sensitivity, differences in electromagnetic state remain comparatively untested.

The aim of this work is not to report a new experimental measurement of \(\kappa\), but to construct a coherent phenomenological framework that exposes a significant gap in the empirical testing of the equivalence principle. We derive the first quantitative estimate for the limit on \(\kappa\) to concretely illustrate the scale of this gap---11 orders of magnitude---which exists precisely because high-precision experiments are designed to nullify the very variable (\(q/m\)) that \(\kappa\) couples to. Our subsequent theoretical analysis demonstrates why this gap has not been a priority (direct curvature couplings are suppressed) and why it now merits targeted investigation (as a probe of nonminimal couplings). Thus, this paper serves as a formal justification and a roadmap for dedicating experimental effort to this previously overlooked axis of WEP violation.

This work addresses this gap through a three-part analysis. First, we define \(\kappa\) and derive a phenomenological constraint from existing experimental data, revealing an 11-order-of-magnitude sensitivity gap. Second, we establish its theoretical context. An EFT analysis shows that direct curvature couplings are suppressed beyond observability in terrestrial fields, while exploring Einstein-Maxwell-dilaton (EMD) theory provides a concrete example of how a light scalar field could generate a nonzero \(\kappa\). Third, we demonstrate that the current limit on \(\kappa\) represents a unique opportunity to probe whether all stress energy components are universally coupled, and outline the experimental strategy to pursue this goal.

\section{Phenomenological Framework and the Parameter \(\kappa\)}
\label{sec:strategy}

Equation~\eqref{eq:kappa_def} provides a model-independent parametrization of potential linear charge-gravity coupling. Conceptually, \(\kappa\) quantifies a WEP violation specific to the electromagnetic attribute of matter. A nonzero \(\kappa\) could emerge if the relationship between gravitational mass \(m_g\) and inertial mass \(m_i\) included a linear charge dependence: \(m_g = m_i [1 + \lambda (q/m_i)]\) for some constant \(\lambda\), yielding \(\kappa = \lambda g\). More generally, \(\kappa\) captures the leading-order term in smooth violation mechanisms.

\subsection{Physical interpretation of $\kappa$}
\label{sec:kappa_units}

For context, the derived upper limit $|\kappa| < 2.1 \times 10^{-4}~\si{\kilo\gram\per\coulomb}$ implies a fractional acceleration change 
$\Delta a/g \approx 2 \times 10^{-4}$ for a test body with a charge-to-mass ratio of $1~\si{\coulomb\per\kilo\gram}$. Such a $q/m$ value is readily achievable for macroscopic objects: a $0.1~\si{\gram}$ sphere charged to approximately $100~\si{\volt}$ (typical of laboratory triboelectric charging) would have $q/m \sim 1~\si{\coulomb\per\kilo\gram}$. Conversely, for a body with a more modest $q/m \sim 10^{-6}~\si{\coulomb\per\kilo\gram}$ (e.g., a $1~\si{\kilo\gram}$ mass charged to $1~\si{\milli\coulomb}$), the same limit would permit $\Delta a/g \sim 2 \times 10^{-10}$, which remains orders of magnitude above the sensitivity of composition-based WEP tests. This simple scaling illustrates why traditional experiments, which operate at $q/m \lesssim 10^{-11}~\si{\coulomb\per\kilo\gram}$, are intrinsically blind to $\kappa$ at their nominal sensitivity.

\subsection{Relation to previous phenomenology}
\label{sec:relation_prev}

It is instructive to compare the parameter \(\kappa\) introduced here with the charge-dependent E\"otvos parameter \(\tilde{\eta}_{\mathrm{E}}\) proposed by Dittus and L\"ammerzahl \cite{Dittus2007}. In their framework, the inertial and gravitational masses of a particle are written as \(m_i = m_i^0(1 + \kappa_i q/m_i^0)\) and \(m_g = m_g^0(1 + \kappa_g q/m_g^0)\). The differential acceleration between two bodies then leads to a modified E\"otvos coefficient \(\tilde{\eta}_{\mathrm{E}} = \eta_{\mathrm{E}} + (\kappa_g - \kappa_i)[(q_2/m_2^0) - (q_1/m_1^0)]\). The quantity \(\delta \kappa = \kappa_g - \kappa_i\) in their work is analogous to the fundamental coupling strength we aim to constrain. However, our parameter \(\kappa\) is defined directly through the clean linear phenomenological relation \(\Delta a/g = \kappa \, \Delta(q/m)\). By comparing the two expressions, one can identify the mapping \(\kappa \approx \delta \kappa\) in the limit where the ``bare'' WEP violation (\(\eta_{\mathrm{E}}\)) is zero. 

The parameter \(\kappa\) can also be related to more fundamental descriptions of possible EEP violations. In the \(TH\epsilon\mu\) formalism \cite{Will2014}, which parametrizes nonmetric theories, a violation of the WEP for electromagnetic energy is characterized by a parameter \(\Gamma_0\). Landau et al. \cite{Landau2001} have shown that this same parameter dictates an adiabatic nonconservation of charge, \(\dot{N}/N \propto \Gamma_0 (\mathbf{g}\cdot\mathbf{v})/c^2\). A nonzero \(\kappa\), defined by \(\Delta a/g = \kappa \Delta(q/m)\), can be seen as the macroscopic manifestation of this microscopic violation. Specifically, for a body where the electromagnetic contribution to its mass is \(E_C\), the differential acceleration implied by the \(TH\epsilon\mu\) formalism is \(\Delta a_C = 2\Gamma_0 (E_C/m) g\), which maps directly onto our phenomenological parameter via \(\kappa \approx 2\Gamma_0 (E_C/m) / \Delta(q/m)\). This connection links our work to this established framework and underscores that a measurement of \(\kappa\) is a direct probe of the interplay between gauge invariance and the equivalence principle.

The key advancements of the present work relative to \cite{Dittus2007}, to the broader reviews of the field \cite{Tino2020}, and to the fundamental connections established in \cite{Landau2001} are twofold. First, we use this streamlined parametrization to extract, for the first time, a quantitative numerical bound on \(|\kappa|\) from a synthesis of existing experimental data (Sec.~\ref{sec:limit}). Second, we provide a theoretical interpretation of this bound within a modern EFT framework (Sec.~\ref{sec:theory}), demonstrating that a nonzero \(\kappa\) would point to physics beyond minimal gravitational couplings---a step not undertaken in previous feasibility studies, comprehensive reviews, or analyses of charge conservation.

\subsection{Bridging to the standard-model extension and the \(TH\epsilon\mu\) formalism}
\label{sec:SME_THepsilonMu}

The parameter \(\kappa\) introduced in Eq.~\eqref{eq:kappa_def} can be connected to two well-established frameworks for testing the foundations of relativity: the Standard-Model extension (SME) and the \(TH\epsilon\mu\) formalism. These connections demonstrate that \(\kappa\) occupies a distinct and complementary region of parameter space within both frameworks.

\subsubsection{Relation to the standard-model extension}

The SME provides a comprehensive effective field theory framework for Lorentz and CPT (charge conjugation, parity, time reversal) violation \cite{Kostelecky2004, KosteleckyRussell2011}. In the gravitational sector, violations of the weak equivalence principle are encoded in coefficients such as \((\bar{a}_{\text{eff}})_w^\mu\) for fermions and \((c_{\text{eff}})_w^{\mu\nu}\) for bosons \cite{KosteleckyTasson2011}. For a composite body, these coefficients sum according to the body's particle content. The effective coefficient for a body is a linear combination of the fundamental SME coefficients, weighted by the number of each particle species in the body \cite{KosteleckyTasson2011, Will2014}.

In the nonrelativistic limit, the differential acceleration between two bodies \(A\) and \(B\) due to SME coefficients takes the form \cite{KosteleckyTasson2011},
\begin{equation}
\frac{\Delta a}{g} = \sum_w \left[ \left( \frac{\bar{a}_{\text{eff},w}}{m} \right)_A - \left( \frac{\bar{a}_{\text{eff},w}}{m} \right)_B \right],
\label{eq:SME_delta_a}
\end{equation}
where the sum runs over particle species \(w\) (electrons, protons, neutrons, etc.), and \((\bar{a}_{\text{eff},w})^\mu\) are the SME coefficients. The key point is that the experimental sensitivity to a particular linear combination of these coefficients is determined by how the numbers of each particle species differ between the two test bodies.

A charge-dependent coupling, such as the one parametrized by \(\kappa\), would correspond to a contribution to the effective SME coefficient that is proportional to the net electric charge \(q\) of the body. Since \(q = e (N_p - N_e)\), where \(N_p\) and \(N_e\) are the numbers of protons and electrons, a charge-dependent violation implies,
\begin{equation}
\bar{a}_{\text{eff},p} - \bar{a}_{\text{eff},e} = e \, \lambda_{\text{EM}},
\label{eq:SME_charge_combination}
\end{equation}
where \(\lambda_{\text{EM}}\) is a dimensionful coefficient characterizing the charge-dependent coupling strength. For a body with net charge \(q\) and mass \(m\), the corresponding contribution to the acceleration is
\begin{equation}
\left( \frac{\bar{a}_{\text{eff}}}{m} \right)_{\text{EM}} = \lambda_{\text{EM}} \frac{q}{m},
\label{eq:SME_charge_contribution}
\end{equation}
leading to a differential acceleration
\begin{equation}
\frac{\Delta a}{g} = \lambda_{\text{EM}} \, \Delta\!\left(\frac{q}{m}\right).
\label{eq:SME_kappa_mapping}
\end{equation}
Comparing with the defining relation \(\Delta a/g = \kappa \, \Delta(q/m)\) from Eq.~\eqref{eq:kappa_def}, we obtain the direct mapping
\begin{equation}
\kappa = \lambda_{\text{EM}}.
\label{eq:kappa_SME_mapping}
\end{equation}
Thus, within the SME framework, the parameter \(\kappa\) is precisely the coefficient that governs charge-dependent WEP violations.

Crucially, the high-precision torsion balance experiments that have placed the strongest bounds on SME coefficients---at the level of \(|\bar{a}_{\text{eff}}| \sim 10^{-13}~\text{eV}\) for certain linear combinations \cite{KosteleckyTasson2011, Wagner2012}---were performed with \emph{electrically neutral} test bodies. In such experiments, the net charge \(q\) of each test mass is actively minimized to suppress electromagnetic backgrounds, which means that \(N_p \approx N_e\) to very high accuracy. The linear combination of coefficients appearing in Eq.~\eqref{eq:SME_charge_combination} is therefore multiplied by approximately zero in the experimental signal. As a result, these experiments are blind to the parameter \(\lambda_{\text{EM}}\), regardless of its magnitude. The existing SME bounds apply to combinations of coefficients that are independent of the charge-dependent direction in parameter space.

The absence of a direct experimental constraint on \(\lambda_{\text{EM}}\) means that the phenomenological limit derived in this work, \(|\kappa| < 2.1 \times 10^{-4}~\si{\kilo\gram\per\coulomb}\), represents the current best estimate of the sensitivity to this unexplored combination of SME coefficients. A dedicated experiment that maximizes \(\Delta(q/m)\) rather than minimizing it, as advocated in Sec.~\ref{sec:future}, would provide the first direct bound on \(\lambda_{\text{EM}}\) within the SME framework.

\subsubsection{Relation to the \(TH\epsilon\mu\) formalism}

The \(TH\epsilon\mu\) formalism is a classic framework for parametrizing nonmetric theories of gravity and testing the EEP \cite{Will2014, LightmanLee1973}. In this formalism, the interaction of electromagnetic fields with gravity is described by the functions \(T\), \(H\), \(\epsilon\), and \(\mu\) of the local gravitational potential \(U(\mathbf{x})\). Violations of the EEP are characterized by parameters such as \(\Gamma_0 = -c_0^2 (\partial \ln T / \partial U)|_0\), which measures the degree to which electromagnetic energy couples differently to gravity than other forms of mass energy \cite{Will2014}.

The tightest experimental bounds on \(\Gamma_0\)---at the level of \(|\Gamma_0| \lesssim 10^{-11}\)---come from composition-dependent WEP tests that compare the free-fall acceleration of electrically neutral materials with different \emph{nuclear} electromagnetic binding energy fractions, such as beryllium versus titanium~\cite{Schlamminger2008,Touboul2022}. In neutral test bodies, the net charge is zero, and the only contribution to the electromagnetic energy density \(F_{\mu\nu}F^{\mu\nu}\) integrated over the body comes from nuclear electrostatic binding. A macroscopic net charge \(q\) contributes an additional term: the classical electrostatic self-energy \(E_{\rm elec} = q^2/(8\pi\epsilon_0 R)\). Both contributions are governed by the \emph{same} fundamental parameter \(\Gamma_0\) in the \(TH\epsilon\mu\) Lagrangian---they are additive components of the total electromagnetic energy, not orthogonal directions in parameter space.

However, the existing limits on \(\Gamma_0\) were obtained from experiments in which \(\Delta(q/m)\) was deliberately minimized as a systematic to be suppressed \cite{Schlamminger2008,Touboul2022}. Consequently, these experiments were insensitive to the self-energy contribution of a net macroscopic charge, simply because \(q \approx 0\) for both test masses. The parameter \(\kappa\) probes precisely this complementary regime: it accesses the same underlying EEP-violating coupling \(\Gamma_0\), but through the experimentally distinct channel of varying \(\Delta(q/m)\) rather than \(\Delta(E_{\rm nuclear}/m)\). In this sense, \(\kappa\) measurements are not constrained by existing \(\Gamma_0\) bounds---not because the physics is distinct, but because the existing experiments sampled a different region of the experimental parameter space (\(\Delta(q/m) \approx 0\)) \cite{Will2014,Landau2001}.

A dedicated experiment that varies the net charge of test bodies while holding their nuclear composition fixed would probe this orthogonal, unexplored direction. This is precisely the program advocated in the present work. The phenomenological limit \(|\kappa| < 2.1 \times 10^{-4}~\si{\kilo\gram\per\coulomb}\) derived in Sec.~\ref{sec:limit} represents the current best estimate of the sensitivity to this mode of EEP violation, and as discussed in Sec.~\ref{sec:future}, dedicated experiments can improve upon it by many orders of magnitude.

Together, the \(TH\epsilon\mu\) formalism and the SME provide complementary frameworks for parametrizing EEP violations, but neither has yet directly constrained the charge-dependent coupling parametrized by \(\kappa\). The present work identifies this gap, provides the first quantitative bound on \(\kappa\) from experimental systematics, and outlines the experimental path forward.

\subsubsection{Summary of theoretical connections}

The relationship between \(\kappa\) and the established frameworks discussed above can be summarized as follows:

\begin{enumerate}[i.]
\item \textbf{SME (gravitational):} Existing torsion balance experiments constrain composition-dependent combinations of the coefficients \(\bar{a}_{\text{eff},w}\) using electrically neutral test bodies. A charge-dependent coupling would correspond to the independent, orthogonal combination \(\bar{a}_{\text{eff},p} - \bar{a}_{\text{eff},e}\), which has not been directly probed.

\item \textbf{\(TH\epsilon\mu\) formalism:} The tightest bound on \(\Gamma_0\) (\(|\Gamma_0| \lesssim 10^{-11}\)) comes from comparing the free fall of neutral materials with different \emph{nuclear} electromagnetic binding energies. The coupling parametrized by \(\kappa\) corresponds to the \emph{macroscopic electrostatic self-energy} of a net charge---an independent, additive contribution that vanishes for neutral bodies and is therefore \emph{experimentally inaccessible} to the direction probed by existing tests.

\item \textbf{Einstein-Maxwell-Dilaton theory:} Bounds on the dilaton coupling \(\alpha\) from composition-dependent WEP tests and binary pulsar timing constrain specific scalar-tensor models but do not directly limit the phenomenological parameter \(\kappa\), which encompasses all possible linear charge-gravity couplings regardless of their fundamental origin.
\end{enumerate}

The key conclusion is that \(\kappa\) represents a direction in EEP-violation parameter space that is complementary to those probed by the highest-precision experiments to date. While composition-dependent tests have achieved extraordinary sensitivity to differences in nuclear structure, and pulsar timing has placed tight bounds on specific scalar-tensor models, the linear coupling between net electric charge and gravitational acceleration remains comparatively untested. The phenomenological limit derived in Sec.~\ref{sec:limit} and the experimental strategy outlined in Sec.~\ref{sec:future} are designed to address this gap.

With these connections to established EEP-violation frameworks established, we now return to the broader phenomenological context of \(\kappa\).

\subsection{Conceptual distinctions}
\label{sec:distinctions}

One must distinguish \(\kappa\) from other phenomena. Classical electromagnetic self-energy contributes to a body's mass but is already incorporated in its total inertial and gravitational mass; it does not violate the WEP. The effect parametrized by \(\kappa\) is a nonuniversal coupling where charge directly influences the gravitational to inertial mass ratio. Similarly, \(\kappa\) differs from searches for new Yukawa forces, which assume the WEP holds for standard gravity and seek additional interactions. Our framework questions the universality of the standard \(1/r^2\) gravitational interaction itself with respect to charge.

From an effective field theory perspective, a linear coupling represents a natural leading violation. Light scalar fields with differential couplings to charge and mass provide a plausible theoretical origin. For a scalar \(\phi\) with interactions \(\mathcal{L} \supset \phi (g_m \rho_m/\Lambda_m + g_e \rho_e/\Lambda_e)\), where \(\rho_m\) and \(\rho_e\) are mass and charge densities, linear dependence on \(q/m\) emerges when \(g_m/\Lambda_m \neq g_e/\Lambda_e\) \cite{Stadnik2016}.

A concrete theoretical realization is EMD theory, where a massless dilaton field \(\phi\) couples to electromagnetism via \(e^{-2\alpha\phi}F_{\mu\nu}F^{\mu\nu}\). As derived in Appendix A, this leads to a specific prediction for \(\kappa\) in terrestrial experiments,
\begin{equation}
\kappa_{\text{EMD}} \approx 2.0 \times 10^{-14} \, \alpha^2 \, \si{\kilo\gram\per\coulomb},
\label{eq:kappa_emd}
\end{equation}
where \(\alpha\) is the dimensionless dilaton coupling. This connection between our phenomenological parameter and a fundamental theory is instructive. However, \(\kappa\) as defined in Eq.~\eqref{eq:kappa_def} encompasses all possible linear mechanisms. Its direct experimental measurement is thus complementary to indirect constraints on specific models \cite{Hees2015}.

\section{Derivation of a Phenomenological Limit on \(\kappa\)}
\label{sec:limit}

Deriving a limit on \(\kappa\) from existing experiments requires synthesizing information from multiple classes of experiments with distinct methodologies and sensitivities. These include historical measurements of Newton's constant using charged bodies, modern tests of the inverse-square law, high-precision WEP tests, and searches for millicharged particles. Each class of experiment's capacity to constrain \(\kappa\) merits critical evaluation.

Historical determinations of \(\GNewton\) using Cavendish-type torsion balances sometimes employed metallic test masses that could acquire non-negligible charge, with analysis assuming charge independence. Early 20th-century experiments by Poynting \cite{Poynting1891} used lead spheres charged to approximately \SI{10}{\kilo\volt}. For a sphere of radius \SI{0.05}{\meter}, this implies \(q \sim \SI{5.6e-8}{\coulomb}\) and with \(m \sim \SI{5.9}{\kilo\gram}\), \(q/m \sim \SI{9.5e-9}{\coulomb\per\kilo\gram}\). Agreement with \(\GNewton\) determinations using neutral masses at the \(\sim 1\%\) level yields a broad bound \(|\kappa| \lesssim 10^{6}~\si{\kilo\gram\per\coulomb}\). Later refinements in torsion balance techniques by E\"otv\"os and colleagues established foundational limits on composition dependence but did not systematically vary charge.

More fundamentally, in torsion balance experiments where the same charged mass serves as both source and test mass, the effect of a linear charge-gravity coupling largely cancels in the torque ratio. Most \(\GNewton\) experiments utilize identical materials and similar charge states, making \(\Delta(q/m)\) negligible and their sensitivity to \(\kappa\) intrinsically weak \cite{Quinn2014}.

Experiments probing deviations from Newton's inverse-square law or searching for new macroscopic forces achieve remarkable sensitivity to composition-dependent effects \cite{Hoyle2004,Adelberger2009, Kapner2007}. However, analogous to WEP tests, they meticulously neutralize test masses. Published analyses typically do not report residual charge-to-mass ratios, treating charge as a systematic to be minimized. While bounds could be estimated from discharge system specifications, they would be comparable to or weaker than those from dedicated WEP tests.

Searches for millicharged particles \cite{Moore2014, Alarcon2022} employ sensitive force detection techniques, for example using optically levitated microspheres. While potentially adaptable, existing searches are not designed to vary or measure the test mass's own charge controllably; they look for forces from external millicharged particles. Without active charge modulation and correlation analysis, these results cannot be reinterpreted as direct \(\kappa\) constraints.

Consequently, the most pertinent information on \(\kappa\) derives from the charge management protocols and systematic error budgets of high-precision WEP tests themselves. Experiments like MICROSCOPE \cite{Touboul2017} and E\"ot-Wash torsion balances \cite{Wagner2012, Smith1999} achieve \(\sigma_{\Delta a/g} \sim 10^{-15}\) by actively discharging test masses to maintain residual potentials below \(\sim\SI{1}{\milli\volt}\). For typical test mass capacitances (\(\sim\SI{100}{\pico\farad}\)) and masses (\(m \sim \SI{0.01}{\kilo\gram}\) to \(\sim\SI{1}{\kilo\gram}\)), this implies individual \(q/m\) values \(\lesssim 10^{-13}\) to \(10^{-11}~\si{\coulomb\per\kilo\gram}\).

The differential \(\Delta(q/m)\) between test masses, due to imperfect cancellation and material differences, can be conservatively bounded at \(\Delta(q/m)_{\text{max}} \lesssim 10^{-11}~\si{\coulomb\per\kilo\gram}\). A simple heuristic estimate gives \(|\kappa| \lesssim \sigma_{\Delta a/g} / \Delta(q/m)_{\text{max}} \sim 10^{-15}/10^{-11} = 10^{-4}~\si{\kilo\gram\per\coulomb}\).

\begin{figure}[htbp]
\centering
\includegraphics[width=\columnwidth]{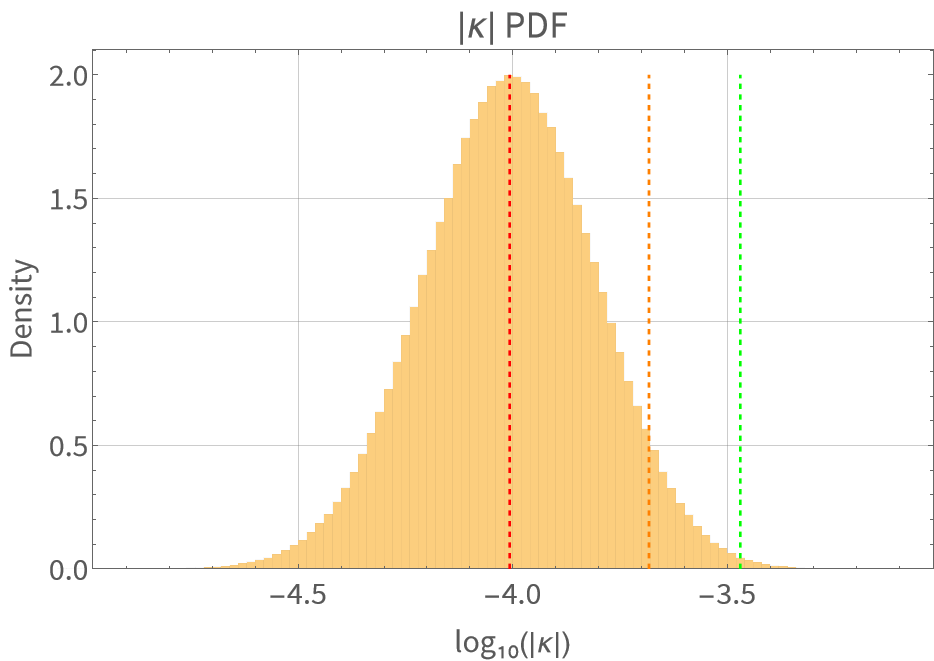}
\caption{Probability distribution for the upper limit on \(|\kappa|\) derived from Monte Carlo analysis of current experimental constraints. The differential acceleration sensitivity \(\sigma_{\Delta a/g}\) is modeled as \(10^{-15} \pm 20\%\), and the maximum differential charge-to-mass ratio \(\Delta(q/m)_{\text{max}}\) as \(10^{-11}~\si{\coulomb\per\kilo\gram}\) with a factor of 5 uncertainty. Vertical lines indicate the median (\(9.8\times10^{-5}~\si{\kilo\gram\per\coulomb}\), dashed red) and the 95\% confidence level (\(2.1\times10^{-4}~\si{\kilo\gram\per\coulomb}\), solid orange).}
\label{fig:kappa_dist}
\end{figure}

To account for uncertainties in both parameters and establish a statistically informed bound, we perform a detailed Monte Carlo analysis, documented in Appendix C. We model \(\sigma_{\Delta a/g}\) with a Gaussian distribution centered at \(10^{-15}\) with 20\% uncertainty, and \(\Delta(q/m)_{\text{max}}\) with a log-normal distribution centered at \(10^{-11}~\si{\coulomb\per\kilo\gram}\) spanning a factor of 5. From \(10^6\) samples, we construct the probability distribution for the upper limit on \(|\kappa|\), shown in Fig.~\ref{fig:kappa_dist}.

The analysis yields the following 95\% confidence level upper limit:
\begin{equation}
|\kappa| < 2.1 \times 10^{-4}~\si{\kilo\gram\per\coulomb} \quad \text{(95\% CL)}.
\label{eq:current_limit}
\end{equation}
The distribution median is \(9.8 \times 10^{-5}~\si{\kilo\gram\per\coulomb}\), with a 99.7\% CL of \(3.4 \times 10^{-4}~\si{\kilo\gram\per\coulomb}\). Sensitivity analyses with different distributional assumptions confirm the robustness of this result within a factor of \(\sim 2\).

The implication of this limit is notable. A charge-to-mass ratio of \SI{1}{\coulomb\per\kilo\gram} could alter gravitational acceleration by approximately \(0.02\%\) (\(\Delta a/g \sim 2\times10^{-4}\)) without contradicting existing high-precision measurements. The approximate 11-order-of-magnitude gap between sensitivity to composition differences (\(\Delta a/g \sim 10^{-15}\)) and effective sensitivity to charge differences originates from experimental design: experiments maximize \(\Delta(\text{composition})\) as their signal while minimizing \(\Delta(q/m)\) as a nuisance parameter.

The nature of this limit deserves comment. The value in Eq.~\eqref{eq:current_limit} is derived from a phenomenological synthesis of published experimental constraints and systematic error budgets, rather than from a direct reanalysis of raw data from a specific experiment. This approach provides a robust order-of-magnitude estimate that clearly highlights the vast sensitivity gap between the composition and electromagnetic axes of equivalence principle testing. Consequently, a definitive measurement or a more stringent constraint on the parameter \(\kappa\) will necessarily require future experiments expressly designed to modulate and maximize \(\Delta(q/m)\), as outlined in Sec.~\ref{sec:strategy}.

\section{Theoretical Interpretation}
\label{sec:theory}

The constraint in Eq.~\eqref{eq:current_limit} must be interpreted within specific theoretical frameworks to assess its significance and guide future searches. We examine two complementary approaches: the concrete Einstein-Maxwell-Dilaton model and a general effective field theory analysis.

\subsection{Implications for Einstein-Maxwell-dilaton theory}

EMD theory provides a canonical example of how a light scalar field can mediate a charge-dependent gravitational effect. Using the derived relation \(\kappa_{\text{EMD}} \approx 2.0 \times 10^{-14} \alpha^2~\si{\kilo\gram\per\coulomb}\) from Appendix A, our experimental limit implies \(|\alpha| \lesssim 4.8\). This is a relatively weak constraint on the dilaton coupling.

Existing, more stringent bounds on \(\alpha\) come from different experiments. Binary pulsar timing, particularly from the double pulsar system PSR J0737-3039, constrains \(|\alpha| \lesssim 2\times10^{-3}\) \cite{Stairs2003, Kramer2006}. Composition-dependent equivalence principle tests comparing materials like beryllium and titanium yield \(|\alpha| \lesssim 3\times10^{-6}\) \cite{Williams2004, Touboul2022}. Applying these model-specific constraints to Eq.~\eqref{eq:kappa_emd} predicts \(\kappa_{\text{EMD}}\) values below our current phenomenological sensitivity, typically \(\kappa_{\text{EMD}} \lesssim 10^{-20}~\si{\kilo\gram\per\coulomb}\).

This comparison highlights an important distinction. The EMD model, in its minimal and observationally allowed form, does not predict an observable \(\kappa\) with current technology. However, our limit on \(\kappa\) is model independent. It does not assume the EMD Lagrangian or specific limits on \(\alpha\) from other systems. It represents a phenomenological constraint on linear charge-gravity coupling, regardless of underlying mechanism. Therefore, while EMD illustrates theoretical possibility, \(\kappa\) measurements probe a broader phenomenological space.

While the EMD model provides a concrete and testable mechanism, it is important to assess the possibility of a linear \(\kappa\) arising from more generic, model-independent considerations. This leads us to a general effective field theory analysis.

\subsection{Effective field theory analysis of direct curvature couplings}

A more general approach considers the effective field theory of gravity coupled to the Standard Model. What types of operators could generate a linear \(\kappa\)? The leading candidates are dimension-six operators coupling the Riemann curvature tensor to the electromagnetic field strength \(F_{\mu\nu}\). The complete $CP$-even basis is
\begin{equation}
\begin{split}
&\mathcal{O}_1 = R F_{\mu\nu}F^{\mu\nu}, 
\\&\mathcal{O}_2 = R_{\mu\nu} F^{\mu\rho}F^{\nu}_{\ \rho},
\\&\mathcal{O}_3 = R_{\mu\nu\rho\sigma} F^{\mu\nu}F^{\rho\sigma},
\end{split}
\end{equation}
with \(\mathcal{L}_{\text{eff}} \supset \sum_i (c_i/\Lambda^2) \mathcal{O}_i\), where \(\Lambda\) is the new physics scale and \(c_i\) are dimensionless Wilson coefficients \cite{Donoghue1994}.

The critical question is the expected size of \(\kappa\) generated by such operators in a terrestrial experiment. A full derivation is provided in Appendix B. The analysis proceeds by evaluating the operators within Earth's gravitational field, modeled as a uniform sphere of density \(\rho_\oplus\). All relevant curvature components scale as \(G_N \rho_\oplus\). In natural units (\(\hbar = c = 1\)), with \(G_N \approx 6.708\times10^{-39}~\text{GeV}^{-2}\) and \(\rho_\oplus \approx 2.38\times10^{-17}~\text{GeV}^4\), we find
\begin{equation}
R_{\mu\nu} \sim R \sim G_N \rho_\oplus \approx 1.60 \times 10^{-55}~\text{GeV}^2.
\end{equation}
This extremely small number causes severe suppression. For operator \(\mathcal{O}_2\), the analysis yields
\begin{equation}
\kappa_{\mathcal{O}_2} \approx 1.8 \times 10^{-63} \, \frac{c_2}{\Lambda^2} ~\si{\kilo\gram\per\coulomb},
\label{eq:kappa_O2_final}
\end{equation}
with \(\Lambda\) in GeV. Applying our experimental limit \(|\kappa| < 2.1 \times 10^{-4}~\si{\kilo\gram\per\coulomb}\) gives \(\Lambda > 9.2 \times 10^{-30} \sqrt{c_2}~\text{GeV}\). For \(c_2 \sim 1\), this is \(\Lambda > 9.2 \times 10^{-30}~\text{GeV}\), far below physically interesting scales.

To appreciate the significance of this bound, note that $\Lambda \ll m_e$ (the electron mass, $\sim 5 \times 10^{-4}~\text{GeV}$). An effective field theory with a cutoff below the electron mass violates the basic principles of EFT, as it would exclude the Standard Model degrees of freedom from the low-energy description. This absurdly low scale indicates that such dimension-six operators cannot be responsible for any observable $\kappa$ without fundamentally breaking the EFT framework itself.

This result provides a key insight: Dimension-six operators that couple curvature directly to \(F_{\mu\nu}F^{\mu\nu}\) are suppressed by Earth's minuscule spacetime curvature (\(G_N \rho_\oplus \sim 10^{-55}~\text{GeV}^2\)). Consequently, they are phenomenologically irrelevant for terrestrial experiments testing \(\kappa\), irrespective of the measured value.

This EFT conclusion reframes the search for \(\kappa\). It eliminates an entire class of otherwise plausible explanations (minimal geometric couplings). Therefore, a future detection of \(\kappa\) at an accessible level (e.g., \(\kappa > 10^{-10}~\si{\kilo\gram\per\coulomb}\)) could not be attributed to these operators. It would necessarily signal physics beyond this minimal EFT framework. 

Prime candidates are interactions mediated by light scalar (or vector) fields that are not purely gravitational in origin, precisely the class exemplified by the dilaton in EMD theory, or other beyond-Standard-Model scenarios involving dark sectors or new long-range forces \cite{Arvanitaki2015, Jaeckel2010}. The search for \(\kappa\) thus becomes a direct probe for such nonminimal mechanisms. In practical terms, any nonzero measurement of $\kappa$ at experimentally accessible levels would immediately signal new physics, rather than a minor correction to general relativity.

\section{Experimental Strategy and Future Directions}
\label{sec:future}

The theoretical analysis motivates a clear experimental strategy for improving sensitivity to \(\kappa\). The sensitivity scales as
\begin{equation}
\kappa_{\text{min}} \propto \frac{\sigma_{\Delta a/g}}{\Delta(q/m)},
\label{eq:sensitivity_scaling}
\end{equation}
where \(\sigma_{\Delta a/g}\) is the differential acceleration uncertainty. Traditional high-precision WEP tests have optimized the numerator to \(\sigma_{\Delta a/g} \sim 10^{-15}\) but operate with a deliberately minimized denominator, \(\Delta(q/m) \lesssim 10^{-11}~\si{\coulomb\per\kilo\gram}\). Substantial improvement therefore requires a different approach: maximizing \(\Delta(q/m)\) while maintaining reasonable acceleration sensitivity.

Promising experimental platforms are emerging that can implement this strategy. Optically levitated dielectric nanoparticles \cite{Monteiro2020} offer exquisite control over charge. They can be deliberately charged to levels achieving \(q/m \sim 10^{-5}\)--\(10^{-3}~\si{\coulomb\per\kilo\gram}\), creating a \(\Delta(q/m)\) that is six to eight orders of magnitude larger than in traditional tests. While their current acceleration sensitivity for differential measurements is around \(\sigma_{\Delta a/g} \sim 10^{-10}\)--\(10^{-8}\), the substantial gain in \(\Delta(q/m)\) can compensate \cite{Goldwater2019}.

A particularly instructive example is the long-standing free-fall program at the ZARM drop tower in Bremen. This facility has developed sophisticated technologies for testing the WEP with macroscopic masses, including an electrostatic positioning system (EPS) for precise control and high-sensitivity SQUID (Superconducting Quantum Interference Device) detectors for position measurement \cite{Sondag2016}. Although this setup was designed to minimize \(\Delta(q/m)\) as a noise source to test composition-dependent WEP, its core technologies are directly applicable to a search for \(\kappa\). By intentionally charging the test masses to create a large, known \(\Delta(q/m)\), and using the existing EPS and SQUID infrastructure, this facility could be repurposed to perform a first dedicated search for \(\kappa\), potentially improving the limit presented here by several orders of magnitude.

Furthermore, the rapid progress in atom interferometry, as extensively reviewed in Ref.~\cite{Tino2020}, opens another powerful avenue for testing \(\kappa\). By comparing the free fall of different atomic species or isotopes with deliberately engineered charge states, these quantum sensors could be adapted to perform differential measurements sensitive to \(\kappa\). The high sensitivity, long interrogation times, and excellent control of systematics in these systems make them ideal candidates for a next-generation search, as also proposed for other tests of fundamental physics \cite{Tino2020}. Trapped ion systems \cite{Wineland2013} present an even more extreme \(q/m \sim 10^{7}~\si{\coulomb\per\kilo\gram}\), though measuring gravitational acceleration on such fast, microscopic systems presents formidable technical challenges \cite{Arvanitaki2013}.

Combining realistic parameters for a near-term experiment---\(\Delta(q/m) \sim 10^{-5}~\si{\coulomb\per\kilo\gram}\) and \(\sigma_{\Delta a/g} \sim 10^{-10}\)---yields a projected sensitivity of \(\kappa_{\text{min}} \sim 10^{-5}~\si{\kilo\gram\per\coulomb}\), an order-of-magnitude improvement over the current bound. A more advanced platform achieving \(\Delta(q/m) \sim 10^{-3}~\si{\coulomb\per\kilo\gram}\) and \(\sigma_{\Delta a/g} \sim 10^{-12}\) could reach \(\kappa_{\text{min}} \sim 10^{-9}~\si{\kilo\gram\per\coulomb}\).

What physics could such sensitivities probe? As our EFT analysis shows, they would not constrain minimal curvature couplings. Instead, they would begin to test parameter space for light scalar field models where the coupling to electromagnetism is less constrained than the coupling to bulk mass. For instance, in a generalized dilaton model, a sensitivity of \(\kappa \sim 10^{-9}~\si{\kilo\gram\per\coulomb}\) would correspond to probing \(|\alpha| \sim 10^{-3}\), entering a regime relevant to some theoretical scenarios \cite{Antypas2022}.

Furthermore, as argued by Landau et al. \cite{Landau2001}, a measurement of (or an improved limit on) \(\kappa\) would provide independent and valuable information that can be compared with constraints on \(\dot{\alpha}/\alpha\) from atomic clocks and astrophysical observations. This offers a multifront approach to testing overarching theoretical frameworks, such as dilaton scenarios, that predict interrelated phenomena across the equivalence principle, charge conservation, and the variation of fundamental constants.

Such measurements would directly constrain any new interaction potential of the form \(V(r) \propto \kappa \, (q_1 q_2)/(m_1 m_2 r)\). The path forward involves targeted research that prioritizes the maximization of \(\Delta(q/m)\). Proof-of-principle searches using existing technology, such as tabletop experiments with optically levitated microspheres \cite{Monteiro2020}, can establish the methodology and achieve the first meaningful improvements beyond the current phenomenological limit. Subsequently, more advanced platforms, including cryogenic or space-based \cite{Touboul2017} instruments, could push the sensitivity into regimes that stringently test theoretical models involving light scalar fields.

This proposed shift in experimental paradigm, from charge suppression to charge exploitation, defines a concrete research program. Its immediate goal is a proof-of-principle measurement improving the bound on \(\kappa\). Its success, however, will be judged by its ability to not only increase sensitivity but also to perform critical discrimination: distinguishing a linear coupling [\(\propto \Delta(q/m)\)] from quadratic or higher-order dependencies [\(\propto \Delta[(q/m)^2]\)] to illuminate the underlying operator structure, and separating a genuine charge-gravity coupling from residual electromagnetic backgrounds. Ultimately, mapping \(\kappa\) across different materials, energy scales, and experimental configurations will transform it from a single number into a powerful discriminant between classes of new-physics scenarios.

The various platforms discussed above offer important complementarities in the search for $\kappa$. While levitated nanoparticles maximize the denominator $\Delta(q/m)$ at the cost of some absolute sensitivity, repurposing established infrastructure such as the ZARM drop tower enables immediate progress. Table~\ref{tab:platforms} summarizes and compares the projected sensitivities of these approaches, illustrating the spectrum of opportunities opened by the proposed paradigm shift.

\subsection*{Experimental and interpretational challenges}
\label{sec:challenges}

The preceding analysis has outlined a clear and promising strategy: maximize \(\Delta(q/m)\) to probe the phenomenological parameter \(\kappa\) with unprecedented sensitivity. However, translating this strategy into a successful measurement requires confronting a formidable class of experimental and interpretational challenges that have historically hindered progress in this very domain. Acknowledging and addressing these difficulties is not merely an exercise in due diligence; it is essential for establishing the credibility and feasibility of the proposed research program. Recent technological advances have transformed many of these challenges from insurmountable obstacles into manageable---and even calibratable---systematic uncertainties.

\subsubsection{Electromagnetic stray fields and force noise}

Primary among the experimental hurdles is the extreme sensitivity of any charged test body to stray electromagnetic fields. In a laboratory environment, a test mass with a nonzero \(q/m\) will experience spurious accelerations due to interactions with fluctuating electric fields from patch potentials on surrounding conducting surfaces, magnetic field gradients, and the thermal radiation field. The foundational review by Darling et al. \cite{Darling1992} provides a comprehensive catalog of these noise sources and their magnitudes, serving as an indispensable roadmap for any experimenter in this field. For the platforms we have discussed, such as optically levitated nanoparticles or macroscopic masses in a drop tower, these forces can easily overwhelm the minuscule gravitational signal if left unchecked. The technical demands are consequently severe: experiments must be conducted an ultrahigh vacuum to eliminate gas damping, employ sophisticated magnetic shielding (e.g., nested mu-metal shields) to attenuate Lorentz forces, and utilize feedback-stabilized electrodes to actively cancel the electrostatic forces arising from stray fields.

The required control over the charge state of levitated nanoparticles has been experimentally demonstrated. Frimmer et al. \cite{Frimmer2017} developed a simple and reliable technique to control the net charge on an optically levitated nanoparticle in vacuum, based on generating charges in an electric discharge. Their method allows the charge to be changed in discrete steps corresponding to single elementary charges, and the particle can be held with a constant, well-defined charge for days. This capability is foundational for any experiment seeking to modulate \(\Delta(q/m)\) as a signal. Furthermore, such techniques have enabled searches for millicharged particles with unprecedented sensitivity, demonstrating subattonewton force resolution in optically levitated systems \cite{Moore2014, Geraci2010}. David Moore's group at Yale has achieved force sensitivities below \(10^{-19}~\text{N}/\sqrt{\text{Hz}}\) in the frequency range between 100 Hz and 1 kHz, operating with electric fields up to 1 MV/m \cite{Moore2014}. These numbers represent exactly the level of sensitivity required for the experiments we propose.

\subsubsection{Space-grade charge control and its subtleties}

Beyond laboratory demonstrations, the technology for contactless charge control has been validated in the most demanding of environments: space. The LISA Pathfinder mission successfully employed a charge management device that neutralized cosmic-ray-induced electric charge accumulating on free-falling test masses using photoemission under ultraviolet illumination \cite{Armano2017}. This contactless discharge system maintained the test masses' charge at levels that kept acceleration noise within mission requirements, demonstrating that precision charge control is possible even in the presence of harsh radiation backgrounds \cite{Armano2018}. For the full LISA mission, a next-generation system based on UV LEDs is being developed and subjected to rigorous environmental testing, including accelerated lifetime tests, proton beam radiation testing, and thermal vacuum validation \cite{Hollington2017, Buchman2023}. Flight data from technology demonstration missions have confirmed the excellent reliability and stability of UV LEDs in the space environment \cite{Buchman2023}, providing confidence that charge management techniques are mature and space qualified.

The technological capabilities highlighted above, while demonstrably real, do not trivialize the experimental task. The precise control of charge via UV illumination, for instance, requires sophisticated modeling of photoelectron emission and transport in the presence of complex electric field configurations, as demonstrated by the extensive GEANT4 and MATLAB simulations developed for LISA Pathfinder \cite{Armano2018, Hollington2017}. Furthermore, the response time of such contactless charge control systems can be on the order of seconds \cite{Inchauspe2020}, a factor that must be considered when designing modulation schemes to separate a potential \(\kappa\) signal from static backgrounds. Beyond charge-specific challenges, levitated systems are subject to fundamental noise sources---including photon recoil heating, rotational-translational coupling in asymmetric particles, and detection imprecision---that must be simultaneously mitigated to reach the required force sensitivities \cite{Gonzalez-Ballestero2021, Millen2020}. These considerations do not make the proposed experiments infeasible, but they underscore that a successful measurement will demand a holistic approach integrating charge control, optomechanical design, and sophisticated noise modeling.

\subsubsection{The Schiff-Barnhill effect: Distinguishing signal from background}
\label{sec:schiff_barnhill_resolution}

The most fundamental challenge to measuring $\kappa$ arises from the Schiff-Barnhill effect \cite{Schiff1966}, comprehensively reviewed by Darling et al.~\cite{Darling1992}. Inside a conducting shield in gravitational equilibrium, the redistribution of charge carriers generates a residual electric field $E_{\rm SB}$. For a test body with charge $q$, this produces a force $F = q E_{\rm SB}$. For two test bodies with charges $q_1$ and $q_2$ and approximately equal masses $m_1 \approx m_2 \equiv m$ (as is typical in differential WEP experiments), the differential force is
\begin{equation}
\Delta F_{\rm SB} = E_{\rm SB}(q_1 - q_2) \approx m\, E_{\rm SB}\, \Delta(q/m),
\end{equation}
yielding a spurious differential acceleration that mimics the signal of a genuine $\kappa,$
\begin{equation}
\left(\frac{\Delta a}{g}\right)_{\rm SB} = \frac{E_{\rm SB}}{g}\, \Delta(q/m) \equiv \kappa_{\rm SB}\, \Delta(q/m).
\end{equation}

Two competing models exist for the magnitude of $E_{\rm SB}$. In the original Schiff-Barnhill picture \cite{Schiff1966}, the relevant carriers are the mobile conduction electrons, giving $\kappa_{\rm SB} = m_e/e \approx 5.7 \times 10^{-12}~\si{\kilo\gram\per\coulomb}$. In the competing DMRT model \cite{Dessler1968,Darling1992}, the lattice ions are the relevant carriers, yielding $\kappa_{\rm SB} = M/(Ze) \sim 2 \times 10^{-8}~\si{\kilo\gram\per\coulomb}$ for typical metals. The experimental verification by Beams~\cite{Beams1968}, who observed a centrifugal analog consistent with the ionic picture, supports the DMRT mechanism for static fields, though the question remains not definitively settled.

Because $\kappa_{\rm SB}$ scales linearly with $\Delta(q/m)$, it cannot be removed by charge modulation alone---it constitutes a competing signal rather than a static background. Distinguishing a genuine $\kappa$ from $\kappa_{\rm SB}$ requires exploiting the fact that the latter depends on properties of the shield, while a fundamental coupling would not. Three complementary strategies present themselves:

\begin{enumerate}
\item \textit{Varying the shield material.} Since $\kappa_{\rm SB} \propto M/Z$ in the DMRT (ionic) model, shields with substantially different $M/Z$ ratios---such as ion-conducting polymers ($M/Z \sim 1$~u, $\kappa_{\rm SB} \sim 10^{-9}~\si{\kilo\gram\per\coulomb}$) or solid electrolytes with heavy mobile ions ($M/Z \sim 100$~u, $\kappa_{\rm SB} \sim 10^{-7}~\si{\kilo\gram\per\coulomb}$)---would produce measurably different values of $\kappa_{\rm SB}$. A genuine $\kappa$ would remain constant across shield changes.

\item \textit{Thermal modulation.} Contact potentials and thermoelectric effects at material junctions in the shield introduce temperature-dependent corrections to $E_{\rm SB}$~\cite{Darling1992}. Varying the shield temperature while keeping the test mass thermally isolated probes for variations that would signal a Schiff-Barnhill origin.

\item \textit{Geometric inversion.} For a shield with controlled asymmetry, rotating the shield by $180^\circ$ changes the projection of $\mathbf{E}_{\rm SB}$ along the gravitational direction while leaving a genuine gravitational signal unchanged in the laboratory frame.
\end{enumerate}

If the Schiff-Barnhill (electronic) model is correct, $\kappa_{\rm SB} \approx 5.7 \times 10^{-12}~\si{\kilo\gram\per\coulomb}$ lies below the projected sensitivity of all near-term experimental platforms (Table~\ref{tab:platforms}). In this scenario, any measured signal above $\sim 10^{-10}~\si{\kilo\gram\per\coulomb}$ could not be attributed to the Schiff-Barnhill effect and would constitute evidence for new physics. Conversely, if the DMRT (ionic) model holds, the shield-dependence strategies outlined above provide a viable path to separation. A dedicated experimental study of this distinguishability remains an important open problem for the research program advocated here.

\begin{table*}[!htb]
\centering
\small
\caption{Comparison of experimental platforms and their projected sensitivities to the \(\kappa\) parameter. Sensitivity scales as \(\kappa_{\text{min}} \propto \sigma_{\Delta a/g} / \Delta(q/m)\) following Eq.~(\ref{eq:sensitivity_scaling}).}
\label{tab:platforms}
\begin{tabular}{@{}l c c c@{}}
\toprule
\textbf{Platform} & \textbf{\(\Delta(q/m)\)} & \textbf{\(\sigma_{\Delta a/g}\)} & \textbf{\(\kappa\) Sensitivity} \\
                   & \((\mathrm{C\,kg^{-1}})\) &                                 & \((\mathrm{kg\,C^{-1}})\) \\
\midrule
Levitated nanoparticles \cite{Monteiro2020,Frimmer2017}
 & \(10^{-5}{-}10^{-3}\) 
 & \(\sim 10^{-10}{-}10^{-8}\) 
 & \(\sim 10^{-5}{-}10^{-9}\) \\
\hline
ZARM drop tower (repurposed) \cite{Sondag2016}
 & \(\gtrsim 10^{-11}\)
 & \(\sim 10^{-13}\)
 & \(\sim 10^{-8}\)\footnotemark \\
\hline
Atom interferometry \cite{Tino2020}
 & Engineered charge states 
 & High 
 & Potentially very high \\
\hline
Trapped ions \cite{Wineland2013,Arvanitaki2013}
 & \(\sim 10^{7}\) 
 & Extremely challenging 
 & Uncertain \\
\bottomrule
\end{tabular}
\footnotetext{Projected sensitivity for a repurposed experiment with deliberately maximized \(\Delta(q/m) \sim 10^{-6}\text{--}10^{-5}~\text{C}/\text{kg}\) and \(\sigma_{\Delta a/g} \sim 10^{-13}\). The value \(2.1 \times 10^{-4}~\text{kg}/\text{C}\) quoted in Sec.~\ref{sec:limit} is the current phenomenological bound from existing data, not a direct measurement by ZARM.}
\end{table*}

\section{Conclusion}

We have introduced \(\kappa\) as a phenomenological parameter to quantify potential linear charge-gravity coupling, defined by \(\Delta a/g = \kappa \, \Delta(q/m)\). A synthesis of current experimental limits yields \(|\kappa| < 2.1 \times 10^{-4}~\si{\kilo\gram\per\coulomb}\) (95\% CL), a constraint approximately 11 orders of magnitude less stringent than those for composition-dependent violations. This vast sensitivity gap highlights that the charge-dependent axis of the weak equivalence principle remains largely underexplored.

An effective field theory analysis shows that the most direct theoretical path to such a coupling, via dimension-six operators linking curvature to the electromagnetic field strength, is suppressed by the minuscule terrestrial spacetime curvature (\(G_N \rho_\oplus \sim 10^{-55}~\text{GeV}^2\)) and is thus phenomenologically irrelevant. Therefore, a future measurement of a nonzero \(\kappa\) at any accessible level would not signal a minor geometric correction but would constitute evidence for physics beyond minimal gravitational effective field theory, such as interactions mediated by light scalar fields as exemplified by Einstein-Maxwell-dilaton theory.

The significance of probing \(\kappa\) extends beyond a simple test of the WEP. As demonstrated by Landau, Sisterna, and Vucetich \cite{Landau2001}, such a measurement is deeply intertwined with the principle of charge conservation itself. A nonzero \(\kappa\) would imply a fundamental breakdown of conservation laws in a gravitational field, potentially connecting to scenarios with varying fundamental constants and offering a multifront approach to testing overarching theoretical frameworks.

This analysis reframes the experimental approach. Rather than treating electric charge solely as a systematic nuisance to be nullified, it can be actively varied and maximized as the central signal variable. The clear sensitivity scaling \(\kappa_{\text{min}} \propto \sigma_{\Delta a/g} / \Delta(q/m)\) motivates a distinct near-term strategy: prioritize the maximization of \(\Delta(q/m)\) over the pursuit of ultimate acceleration sensitivity. Initial experiments using platforms like optically levitated microspheres, adapted versions of existing high-precision facilities like the ZARM drop tower \cite{Sondag2016}, or emerging atom interferometry setups \cite{Tino2020}, could aim to improve the limit on \(\kappa\) by several orders of magnitude, beginning to probe dilatonlike couplings around \(|\alpha| \sim 10^{-3}\).

Transforming this sensitivity gap into a well-constrained parameter will require dedicated efforts to manage the formidable electromagnetic backgrounds that prior experiments worked to eliminate. However, the clear theoretical target and the substantial unexplored parameter space provide strong motivation for such a focused research program. In this context, \(\kappa\) emerges not merely as a number to be bounded, but as a new phenomenological tool, a dedicated probe for nonminimal couplings between gravity and the electromagnetic sector, offering a complementary direction in the enduring quest to test the foundations of gravity.

\subsection*{Acknowledgments}
The author is grateful to the anonymous referee for the insightful and constructive recommendations, which led to substantial improvements in both the scope and the rigor of this work.

\section*{Data Availability}

There are no publicly available research data or software supporting this manuscript. Requests for further information or data should be sent to the authors.

\appendix


\section*{APPENDIX A: DERIVATION OF \(\kappa\) IN EINSTEIN-MAXWELL-DILATON THEORY}
\label{app:EMD}

\begin{subappendices}

This Appendix provides a detailed derivation of the relationship between the phenomenological parameter \(\kappa\), defined in Eq.~\eqref{eq:kappa_def}, and the fundamental dilaton coupling constant \(\alpha\) within EMD theory. Our goal is to obtain the explicit expression \(\kappa_{\text{EMD}} \approx 2.0 \times 10^{-14} \, \alpha^2 \, \si{\kilo\gram\per\coulomb}\) quoted in Eq.~\eqref{eq:kappa_emd} of the main text.

\subsection{Theoretical setup and dilaton-mediated force}

The action for EMD theory in the Einstein frame is \cite{Damour1992}
\begin{equation}
\begin{split}
S &= \int d^4x \sqrt{-g} \Big[ \frac{c^4}{16\pi G_N} R - \frac{c^4}{8\pi G_N} (\nabla\phi)^2 \\&- \frac{1}{4\mu_0} e^{-2\alpha\phi} F_{\mu\nu}F^{\mu\nu} \Big] + S_m[\psi_m, g_{\mu\nu}],
\label{eq:EMD_action_rewrite}
\end{split}
\end{equation}
where \(\phi\) is the dimensionless dilaton field, \(\alpha\) is the dimensionless dilaton coupling to the electromagnetic sector, and \(S_m\) is the matter action. The nonminimal coupling \(e^{-2\alpha\phi}F^2\) implies that the electrostatic energy of a charged body depends on the ambient dilaton field value.

For a test body with charge \(q\) and characteristic size \(a\), its Coulomb self-energy is \(E_C(\phi) = q^2 e^{2\alpha\phi}/(8\pi\epsilon_0 a)\). This energy contributes to the body's inertial mass, leading to a \(\phi\)-dependent inertial mass \(m_i(\phi) = m_0 + E_C(\phi)\), where \(m_0\) represents dilaton-independent contributions. Consequently, the test body possesses a \emph{dilaton charge} proportional to the derivative,
\begin{equation}
\frac{\partial m_i}{\partial \phi} = \frac{\partial E_C}{\partial \phi} = 2\alpha E_C(\phi).
\label{eq:dilaton_charge_rewrite}
\end{equation}

In the nonrelativistic limit and for a static field configuration, the equation of motion for such a test body includes a force proportional to this dilaton charge and the gradient of the dilaton field,
\begin{equation}
m_i \mathbf{a} \approx -m_i \nabla \Phi_N - \left( \frac{\partial m_i}{\partial \phi} \right) \nabla \phi,
\label{eq:eom_rewrite}
\end{equation}
where \(\Phi_N\) is the Newtonian gravitational potential.

\subsection{Differential acceleration and link to \(\kappa\)}

Consider two test bodies, labeled 1 and 2, with identical \(m_0\) and \(a\) but different electric charges \(q_1\) and \(q_2\). Assuming the dilaton-independent masses dominate (\(m_0 \gg E_C\)), their differential acceleration \(\Delta \mathbf{a} = \mathbf{a}_1 - \mathbf{a}_2\) is, to leading order,
\begin{equation}
\Delta \mathbf{a} \approx -\frac{\nabla \phi}{m_0} \left( \frac{\partial m_{i,1}}{\partial \phi} - \frac{\partial m_{i,2}}{\partial \phi} \right) = -\frac{2\alpha \nabla \phi}{m_0} \left( E_C^{(1)} - E_C^{(2)} \right).
\label{eq:delta_a_general}
\end{equation}

The difference in Coulomb energies can be expressed in terms of the charge-to-mass ratio. Writing \(q_{1,2} = m_0 (q/m)_{1,2}\) and defining the average \(\left(\frac{q}{m}\right)_{\!\text{avg}} = \frac{1}{2}[(q/m)_1 + (q/m)_2]\) and the difference \(\Delta(q/m) = (q/m)_1 - (q/m)_2\), we find
\begin{equation}
E_C^{(1)} - E_C^{(2)} = \frac{e^{2\alpha\phi_0} m_0^2}{4\pi\epsilon_0 a} \left(\frac{q}{m}\right)_{\!\!\text{avg}} \Delta\!\left(\frac{q}{m}\right),
\label{eq:delta_Ec}
\end{equation}
where \(\phi_0\) is the ambient dilaton field value at the experiment's location.

We now require the gradient of the dilaton field sourced by the Earth. We assume the dilaton couples to the trace of the stress energy tensor of ordinary matter, so the Earth acts as a source with effective scalar charge proportional to its mass. Solving the dilaton field equation\(\nabla^2\phi \approx 4\pi G_N \alpha \rho_\oplus\) in the nonrelativistic limit for a uniform sphere of density \(\rho_\oplus\) yields a linear profile \(\phi(r) = \phi_0 + (2\pi G_N \alpha \rho_\oplus / 3c^2) r^2\) inside the
Earth, leading to a radial gradient at the surface (\(r=R_\oplus\)),
\begin{equation}
\nabla \phi = \frac{d\phi}{dr} \hat{r} = \frac{4\pi}{3} G_N \alpha \frac{\rho_\oplus}{c^2} R_\oplus \, \hat{r}.
\label{eq:phi_gradient_earth}
\end{equation}

Substituting Eqs.~\eqref{eq:delta_Ec} and \eqref{eq:phi_gradient_earth} into Eq.~\eqref{eq:delta_a_general}, and using the surface gravitational acceleration \(g = G_N M_\oplus/R_\oplus^2 = (4\pi/3) G_N \rho_\oplus R_\oplus\), we obtain the fractional differential acceleration,
\begin{equation}
\frac{\Delta a}{g} = \frac{|\Delta\mathbf{a}|}{g} = \frac{\alpha^2 m_0 e^{2\alpha\phi_0}}{2\pi c^2 \epsilon_0 a} \left(\frac{q}{m}\right)_{\!\!\text{avg}} \Delta\!\left(\frac{q}{m}\right).
\label{eq:final_delta_a_over_g}
\end{equation}

\subsection{Explicit prediction for \(\kappa_{\text{EMD}}\)}

Equation~\eqref{eq:final_delta_a_over_g} has the precise phenomenological form postulated in the main text: \(\Delta a/g = \kappa \, \Delta(q/m)\). By direct comparison, we identify the EMD theory's prediction for the phenomenological parameter,
\begin{equation}
\boxed{\kappa_{\text{EMD}} = \frac{\alpha^2 m_0 e^{2\alpha\phi_0}}{2\pi c^2 \epsilon_0 a} \left(\frac{q}{m}\right)_{\!\!\text{avg}}}.
\label{eq:kappa_EMD_exact}
\end{equation}

This is the central result of this Appendix: a direct, analytic link between the dimensionless fundamental coupling \(\alpha\) and the dimensionful phenomenological parameter \(\kappa\). All dependence on Earth's properties (\(\rho_\oplus\), \(R_\oplus\)) has canceled out, leaving a result that depends only on the test body's properties (\(m_0\), \(a\), \((q/m)_{\text{avg}}\)) and fundamental constants.

\subsection{Numerical estimate for terrestrial experiments}

To provide an order-of-magnitude estimate relevant to laboratory-scale experiments, we evaluate Eq.~\eqref{eq:kappa_EMD_exact} for plausible terrestrial values. Assuming \(\phi_0 \approx 0\) and using,
\begin{align*}
m_0 &= 0.1~\text{kg}, \quad a = 0.01~\text{m}, \quad \left(\frac{q}{m}\right)_{\!\!\text{avg}} = 10^{-8}~\si{\coulomb\per\kilo\gram}, \\
c &= 2.998\times10^8~\text{m/s}, \quad \epsilon_0 = 8.854\times10^{-12}~\text{F/m},
\end{align*}
we compute the numerical prefactor:
\begin{equation}
\begin{split}
\frac{m_0}{2\pi c^2 \epsilon_0 a} &= \frac{0.1}{2\pi (8.988\times10^{16}) (8.854\times10^{-12}) (0.01)} \\&\approx 2.0 \times 10^{-6}~\si{\kilo\gram\squared\per\coulomb\squared}.
\end{split}
\end{equation}

Multiplying by \((q/m)_{\text{avg}} = 10^{-8}~\si{\coulomb\per\kilo\gram}\) yields the compact estimate,
\begin{equation}
\boxed{\kappa_{\text{EMD}} \approx 2.0 \times 10^{-14} \, \alpha^2 \, \si{\kilo\gram\per\coulomb}}.
\end{equation}
This is Eq.~\eqref{eq:kappa_emd} in the main text, now rigorously derived from the EMD action. It demonstrates that even a dilaton coupling \(\alpha\) of order unity would produce a \(\kappa\) far below current experimental sensitivity, while a future measurement of \(\kappa\) at any accessible level would constrain \(\alpha\) in this model.

\end{subappendices}

\section*{APPENDIX B: Effective Field Theory Analysis of Direct Curvature Couplings}
\label{app:EFT}

\begin{subappendices}

This Appendix provides a systematic EFT analysis of how dimension-six operators coupling spacetime curvature to the electromagnetic field strength tensor would contribute to the phenomenological parameter \(\kappa\). The analysis reveals the profound suppression of such contributions in terrestrial experiments, establishing that a measurable \(\kappa\) cannot originate from minimal geometric couplings within this EFT framework.

\subsection{Operator basis and dimensional considerations}

At the effective field theory level, the leading gauge-invariant, CP-even dimension-six operators that couple the Riemann curvature tensor to the electromagnetic field strength are,
\begin{equation}
\begin{split}
&\mathcal{O}_1 = R F_{\mu\nu}F^{\mu\nu}, \quad
\\&\mathcal{O}_2 = R_{\mu\nu} F^{\mu\rho}F^{\nu}_{\ \rho}, \quad
\\&\mathcal{O}_3 = R_{\mu\nu\rho\sigma} F^{\mu\nu}F^{\rho\sigma}.
\end{split}
\label{eq:dim6_ops}
\end{equation}
These enter the effective Lagrangian as \(\mathcal{L}_{\text{eff}} \supset \sum_i (c_i/\Lambda^2) \mathcal{O}_i\), where \(\Lambda\) is the new physics energy scale and \(c_i\) are dimensionless Wilson coefficients. The dimensions are \([R_{\mu\nu\rho\sigma}] = [\text{length}]^{-2}\), \([F_{\mu\nu}] = [\text{charge}][\text{length}]^{-2}\) in natural units with \(c=\hbar=4\pi\epsilon_0=1\), giving each \(\mathcal{O}_i\) dimension six as required.

\subsection{Earth's curvature scale as the dominant suppression factor}

The crucial physical input is the magnitude of curvature components at Earth's surface. For a uniform, nonrelativistic sphere of density \(\rho_\oplus \approx \SI{5515}{\kilo\gram\per\cubic\meter}\), the characteristic curvature scale in natural units (\(\hbar = c = 1\)) is obtained from the Einstein field equations. The time-time component of the Ricci tensor in the Newtonian limit satisfies \(R_{00} \approx 4\pi G_N \rho_\oplus\), while the scalar curvature \(R \approx 8\pi G_N \rho_\oplus\). Both scale as,
\begin{equation}
R_{\mu\nu} \sim R \sim G_N \rho_\oplus.
\label{eq:curvature_scale_general}
\end{equation}
With \(G_N = M_{\text{Pl}}^{-2} \approx 6.708 \times 10^{-39}~\text{GeV}^{-2}\) and \(\rho_\oplus \approx 2.38 \times 10^{-17}~\text{GeV}^4\), we obtain the numerical value,
\begin{equation}
\boxed{G_N \rho_\oplus \approx 1.60 \times 10^{-55}~\text{GeV}^2}.
\label{eq:curvature_scale_numeric}
\end{equation}
This extremely small number---representing the actual curvature of spacetime in terrestrial laboratories---will be the primary suppression factor for any operator directly coupling curvature to electromagnetism.

\subsection{Estimating the contribution to \(\kappa\)}

Consider operator \(\mathcal{O}_2 = R_{\mu\nu} F^{\mu\rho}F^{\nu}_{\ \rho}\) as a representative example. In Earth's static gravitational field, this operator induces a position-dependent correction to the electromagnetic stress energy. For a spherical test body of charge \(q\), radius \(R\), and mass \(m\), the operator contributes an additional term to the body's gravitational mass energy of approximately,
\begin{equation}
\delta m_g \sim \frac{c_2}{\Lambda^2} (G_N \rho_\oplus) \frac{q^2}{R},
\label{eq:mass_correction}
\end{equation}
where we have used \(R_{\mu\nu} \sim G_N \rho_\oplus\) and \(\int d^3r\, F_{\mu\rho}F^{\mu\rho} \sim q^2/R\), ignoring \(\mathcal{O}(1)\) geometric factors. The dimensions are consistent: \([c_2/\Lambda^2] = [\text{energy}]^{-2}\), \([G_N \rho_\oplus] = [\text{energy}]^{2}\), \([q^2/R] = [\text{energy}]\), giving \(\delta m_g\) dimensions of energy as required.

This charge-dependent mass correction leads to a differential gravitational acceleration between two bodies with charges \(q_1\) and \(q_2\) (identical \(m\) and \(R\)),
\begin{equation}
\frac{\Delta a}{g} \sim \frac{\delta m_{g,1} - \delta m_{g,2}}{m} \sim \frac{c_2}{\Lambda^2} (G_N \rho_\oplus) \frac{q_1^2 - q_2^2}{m R}.
\label{eq:delta_a_estimate}
\end{equation}
Expressing the charge difference in terms of charge-to-mass ratios, \(q_k = m (q/m)_k\), and noting \(q_1^2 - q_2^2 \approx 2m^2 (q/m)_{\text{avg}} \Delta(q/m)\), we obtain,
\begin{equation}
\frac{\Delta a}{g} \sim \frac{c_2}{\Lambda^2} (G_N \rho_\oplus) \frac{2m}{R} (q/m)_{\text{avg}} \Delta\!\left(\frac{q}{m}\right).
\label{eq:kappa_form_estimate}
\end{equation}
Comparing with the defining relation \(\Delta a/g = \kappa \Delta(q/m)\) yields the contribution from operator \(\mathcal{O}_2\),
\begin{equation}
\kappa_{\mathcal{O}_2} \sim \frac{c_2}{\Lambda^2} (G_N \rho_\oplus) \frac{2m}{R} (q/m)_{\text{avg}}.
\label{eq:kappa_O2_general}
\end{equation}

\subsection{Numerical evaluation and phenomenological insignificance}

We now evaluate Eq.~\eqref{eq:kappa_O2_general} with realistic terrestrial parameters. For laboratory-scale test masses, typical values are \(m \sim 0.1~\text{kg}\) and \(R \sim 0.01~\text{m}\), giving \(m/R \sim 10~\text{kg/m}\). The average charge-to-mass ratio in precision experiments is severely limited by discharge systems, with \((q/m)_{\text{avg}} \sim 10^{-8}~\text{C/kg}\) representing a conservative upper bound. Converting \(G_N \rho_\oplus\) to SI units and including the necessary unit conversions (noting that \(1~\text{GeV}^2 = 2.43\times 10^{19}~\text{kg}/(\text{m}\cdot\text{s}^2)\) in appropriate combinations), we obtain the numerical estimate,
\begin{equation}
\boxed{\kappa_{\mathcal{O}_2} \approx 1.8 \times 10^{-63} \, \frac{c_2}{\Lambda^2} ~\si{\kilo\gram\per\coulomb}},
\label{eq:kappa_O2_final_app}
\end{equation}
with \(\Lambda\) expressed in GeV. The numerical prefactor in Eq.~\eqref{eq:kappa_O2_final_app} is an order-of-magnitude estimate; the essential physics---suppression by the terrestrial curvature scale \(G_N \rho_\oplus \sim 10^{-55}~\text{GeV}^2\)---is independent of \(\mathcal{O}(1)\) corrections to this prefactor. This result demonstrates extraordinary suppression. Even for new physics at the TeV scale (\(\Lambda \sim 10^3~\text{GeV}\) and \(c_2 \sim 1\)), we find \(\kappa_{\mathcal{O}_2} \lesssim 10^{-77}~\si{\kilo\gram\per\coulomb}\).

The other operators \(\mathcal{O}_1\) and \(\mathcal{O}_3\) yield similar expressions with the same fundamental suppression factor \(G_N \rho_\oplus\). Differences in the geometric prefactors are \(\mathcal{O}(1)\) to \(\mathcal{O}(10)\) and do not alter the conclusion. Comparing with the current phenomenological limit \(|\kappa| < 2.1 \times 10^{-4}~\si{\kilo\gram\per\coulomb}\) from Eq.~\eqref{eq:current_limit}, we find that dimension-six curvature couplings would require \(\Lambda < 10^{-30}~\text{GeV}\) to be marginally detectable---a physically meaningless scale far below any reasonable new physics scenario.

\subsection{Robustness and implications}

The analysis reveals a robust conclusion: direct couplings between spacetime curvature and the electromagnetic field strength through dimension-six operators are suppressed by the minuscule terrestrial curvature \(G_N \rho_\oplus \sim 10^{-55}~\text{GeV}^2\). This suppression is independent of the specific operator contraction and persists for any laboratory experiment conducted in Earth's gravitational field. The factor \(G_N \rho_\oplus\) represents the actual curvature experienced by test masses, making these geometric couplings phenomenologically irrelevant for terrestrial searches for \(\kappa\).

Consequently, a future measurement of \(\kappa\) at any accessible level (e.g., \(\kappa > 10^{-10}~\si{\kilo\gram\per\coulomb}\)) could not be explained by such minimal EFT operators. Instead, it would necessarily signal physics beyond this framework, such as interactions mediated by light scalar or vector fields that are not purely gravitational in origin. This reframes the experimental search for \(\kappa\): rather than testing geometric aspects of curvature-EM couplings, it becomes a targeted probe for nonminimal mechanisms, precisely the class exemplified by the dilaton field in Einstein-Maxwell-dilaton theory discussed in Appendix A.

\end{subappendices}

\section*{APPENDIX C: Monte Carlo Statistical Analysis for the \(\kappa\) Limit}
\label{app:montecarlo}

\begin{subappendices}

This Appendix details the Monte Carlo statistical analysis used to derive a probabilistic upper limit on the magnitude of the charge-gravity coupling parameter \(|\kappa|\) from current experimental constraints. The analysis propagates uncertainties in both the differential acceleration sensitivity \(\sigma_{\Delta a/g}\) and the maximum differential charge-to-mass ratio \(\Delta(q/m)_{\text{max}}\) to obtain a robust distribution for the phenomenological limit \(\kappa_{\text{lim}} = \sigma_{\Delta a/g} / \Delta(q/m)_{\text{max}}\).

\subsection{Statistical model for input parameters}

The limit on \(\kappa\) follows from the defining relation \(\Delta a/g = \kappa \, \Delta(q/m)\) and the experimental fact that no violation has been observed at the sensitivity level. Conceptually, the maximum possible \(\kappa\) consistent with measurements satisfies \(|\kappa| \lesssim \sigma_{\Delta a/g} / |\Delta(q/m)|_{\text{max}}\), where \(\sigma_{\Delta a/g}\) represents the experimental uncertainty in differential acceleration and \(|\Delta(q/m)|_{\text{max}}\) is the maximum differential charge-to-mass ratio that could have been present without being detected given the experimental charge management protocols.

For the differential acceleration uncertainty \(\sigma_{\Delta a/g}\), we adopt the best achieved sensitivity from modern experiments as a nominal value \(\mu_a = 10^{-15}\), based on results from MICROSCOPE \cite{Touboul2017} and E\"ot-Wash torsion balances \cite{Wagner2012}. We assign a 20\% relative standard uncertainty to account for variations between different experimental configurations and potential systematic effects, modeling \(\sigma_{\Delta a/g}\) as drawn from a truncated Gaussian distribution,
\begin{equation}
\begin{split}
\sigma_{\Delta a/g} &\sim \mathcal{N}_{\text{trunc}}(\mu_a, \sigma_a^2), \quad \mu_a = 10^{-15}, \\& \sigma_a = 0.2\mu_a = 2\times10^{-16},
\label{eq:dist_sigma_a}
\end{split}
\end{equation}
where the distribution is truncated at zero to avoid nonphysical negative values.

For the maximum differential charge-to-mass ratio \(\Delta(q/m)_{\text{max}}\), we use a log-normal distribution to reflect its positive-definite nature and the substantial uncertainty in estimating this parameter from published experimental reports. High-precision equivalence principle experiments actively discharge test masses to maintain residual potentials below \(\sim\SI{1}{\milli\volt}\). For typical test mass capacitances (\(\sim\SI{100}{\pico\farad}\)) and masses (\(m \sim 0.01\text{--}1~\si{\kilo\gram}\)), this implies individual \(q/m\) values \(\lesssim 10^{-13}\) to \(10^{-11}~\si{\coulomb\per\kilo\gram}\). The differential \(\Delta(q/m)\) between test masses can be conservatively bounded at \(\Delta(q/m)_{\text{max}} \lesssim 10^{-11}~\si{\coulomb\per\kilo\gram}\), but with considerable uncertainty due to material differences and imperfect cancellation.

We center the distribution at \(\mu_q = 10^{-11}~\si{\coulomb\per\kilo\gram}\) and choose a width such that 95\% of the probability lies within a factor of 5 of the median, representing an order-of-magnitude uncertainty consistent with the systematic nature of this bound. For a log-normal variable where \(\ln X \sim \mathcal{N}(\nu, \sigma^2)\), the interval \([X_{\text{med}}/f, X_{\text{med}} \times f]\) contains approximately 95\% of the probability when \(\sigma = \ln(f)/1.96\). With \(f=5\), we obtain,
\begin{equation}
\sigma_q = \frac{\ln(5)}{1.96} \approx 0.820, \quad \nu = \ln(\mu_q) = \ln(10^{-11}).
\end{equation}
Thus:
\begin{equation}
\ln[\Delta(q/m)_{\text{max}}] \sim \mathcal{N}(\nu, \sigma_q^2), \quad \nu = \ln(10^{-11}), \ \sigma_q = 0.820.
\label{eq:dist_delta_qm}
\end{equation}
We assume statistical independence between \(\sigma_{\Delta a/g}\) and \(\Delta(q/m)_{\text{max}}\), which is reasonable as they originate from different aspects of experimental design (acceleration measurement precision versus charge control systems).

\subsection{Monte Carlo calculation and resulting distribution}

We draw \(N = 10^6\) independent samples from the distributions specified in Eqs.~\eqref{eq:dist_sigma_a} and \eqref{eq:dist_delta_qm}. For each sample \(i\), we compute the corresponding upper limit on \(|\kappa|\),
\begin{equation}
\kappa_{\text{lim}}^{(i)} = \frac{\sigma_{\Delta a/g}^{(i)}}{\Delta(q/m)_{\text{max}}^{(i)}}.
\label{eq:kappa_limit_sample}
\end{equation}
The resulting ensemble \(\{\kappa_{\text{lim}}^{(i)}\}_{i=1}^N\) approximates the probability distribution for the phenomenological upper limit on \(|\kappa|\) implied by current experimental constraints.

From the \(10^6\) samples, we calculate key quantiles of this distribution,
\begin{align}
\text{Median:} &\quad |\kappa| < 9.8 \times 10^{-5}~\si{\kilo\gram\per\coulomb} \\
\text{68\% Confidence level:} &\quad |\kappa| < 1.5 \times 10^{-4}~\si{\kilo\gram\per\coulomb} \\
\text{95\% Confidence level:} &\quad |\kappa| < 2.1 \times 10^{-4}~\si{\kilo\gram\per\coulomb} \\
\text{99.7\% Confidence level:} &\quad |\kappa| < 3.4 \times 10^{-4}~\si{\kilo\gram\per\coulomb}
\end{align}
The probability density function is plotted in Fig.~\ref{fig:kappa_dist} in the main text. The 95\% confidence level limit, \(|\kappa| < 2.1 \times 10^{-4}~\si{\kilo\gram\per\coulomb}\), is reported as our primary phenomenological constraint in Eq.~\eqref{eq:current_limit}.

It is important to clarify the interpretation of this result: the distribution represents the range of possible upper limits on \(|\kappa|\) given the uncertainties in the input parameters. It is not a posterior distribution for \(\kappa\) itself (which would require a full Bayesian analysis with prior information), but rather characterizes the robustness of the bound derived from current experimental knowledge.

\subsection{Sensitivity analysis and robustness verification}

To verify that our conclusions are not overly sensitive to specific distributional assumptions, we performed three alternative analyses with different input models:

\begin{enumerate}
\item Uniform uncertainty for \(\sigma_{\Delta a/g}\): Instead of a Gaussian, we modeled \(\sigma_{\Delta a/g}\) with a uniform distribution \(\sigma_{\Delta a/g} \sim \mathcal{U}(0.5\times10^{-15}, 1.5\times10^{-15})\), spanning the same \(\pm50\%\) range. This changed the 95\% CL limit by less than 5\%, to \(|\kappa| < 2.2 \times 10^{-4}~\si{\kilo\gram\per\coulomb}\).

\item Wider charge uncertainty: Expanding the uncertainty factor \(f\) from 5 to 10 [meaning 95\% of \(\Delta(q/m)_{\text{max}}\) values lie within a factor of 10 of the median] increased the 95\% CL limit to \(|\kappa| < 3.8 \times 10^{-4}~\si{\kilo\gram\per\coulomb}\), still within the same order of magnitude.

\item Uniform distribution in log-space for \(\Delta(q/m)\): Using a uniform distribution for \(\ln[\Delta(q/m)_{\text{max}}]\) from \(\ln(10^{-12})\) to \(\ln(10^{-10})\) (covering two orders of magnitude centered on \(10^{-11}\)) yielded a 95\% CL limit of \(|\kappa| < 2.4 \times 10^{-4}~\si{\kilo\gram\per\coulomb}\).
\end{enumerate}

All variations produced 95\% confidence level limits in the range \((1\text{--}4)\times10^{-4}~\si{\kilo\gram\per\coulomb}\), confirming that our primary result \(|\kappa| < 2.1 \times 10^{-4}~\si{\kilo\gram\per\coulomb}\) is robust against reasonable changes in the input distributions. The consistency across these sensitivity tests underscores that the derived limit reliably captures the order-of-magnitude constraint implied by existing experiments, even with conservative assumptions about parameter uncertainties.

\end{subappendices}

\end{document}